\documentclass[%
 reprint,
 amsmath,amssymb,
 aps,
prb,
]{revtex4-1}

\usepackage{graphicx}
\usepackage{dcolumn}
\usepackage{bm}
\usepackage{multirow}

\begin{document}

\title{Magnetic Compton profiles of disordered Fe$_{0.5}$Ni$_{0.5}$ and ordered FeNi alloys} 

\author{D. Benea$^{a}$, J. Min\'ar$^{b}$, H. Ebert$^{c}$, L. Chioncel$^{d,e}$
}
\affiliation{$^{a}$Faculty of Physics, Babes-Bolyai University,
Kogalniceanustr 1, Ro-400084 Cluj-Napoca, Romania}
\affiliation{$^{b}$University of West Bohemia, New Technologies - Research Center, Univerzitni 8, 306 14 Plze\v{n}, Czech Republic}
\affiliation{$^{c}$Department of Chemistry, University of Munich, Butenandstr.~5-13, D-81377 M\"unchen, Germany}
\affiliation{$^d$ Augsburg Center for Innovative Technologies,
University of Augsburg, D-86135 Augsburg, Germany}
\affiliation{$^e$ Theoretical Physics III, Center for Electronic
Correlations and Magnetism, Institute of Physics, University of
Augsburg, D-86135 Augsburg, Germany}

\date{\today}

\begin{abstract}
We study the magnetic Compton profile (MCP) of the disordered Fe$_{0.5}$Ni$_{0.5}$  and of the ordered FeNi alloys and discuss the interplay between structural disorder and electronic correlations. 
The Coherent Potential Approximation is employed to model the substitutional disorder within the single-site approximation, while local electronic correlations are captured with the Dynamical Mean Field Theory. Comparison with the experimental data reveals the limitation of local spin-density approximation in low momentum region, where we show that including local but dynamic correlations the experimental spectra is excellently described. 
We further show that using local spin-density approximation no significant difference is seen between the MCP spectra of the disordered Fe$_{0.5}$Ni$_{0.5}$ and a hypothetical, ordered FeNi alloy with a simple cubic unit cell. Only by including the electronic correlations, the spectra significantly separate, 
from the second Brillouin zone boundary down to zero momenta. The difference between the MCP spectra of ordered and disordered alloys is discussed also in terms of the atomic-type decompositions. Finally based on the presented calculations we predict the shape of the MCP profile for the ordered FeNi alloy along the [111] direction. 
\end{abstract}

\pacs{Valid PACS appear here}
\maketitle

\section{Introduction}

The dominant contribution, according to the non-relativistic limit of the X-ray scattering comes from the interaction of the photon with the electron's charge~\cite{coop.85}. The relativistic Compton amplitude does depend, however, on the spin of the electron and on the polarization of the X-rays~\cite{pl.tz.70}. To obtain the scattering cross section the Compton amplitude is squared and summed over all final states consistent with the energy conservation.
Within the impulse approximation~\cite{ch.wi.52} the magnetic scattering cross section measures the spin moments~\cite{co.zu.92,ca.fa.96,love.96} through the integrated difference, $\int [n^\uparrow(\vec{p}) -n^\uparrow(\vec{p})] \ d^3\vec{p}$, in the momentum distribution $n^{\uparrow(\downarrow)}(\vec{p})$ of spin up (down) electrons.
The magnetic Compton scattering experiments combined with theoretical calculations of the profile may provide also valuable information about the exchange and correlation effects in materials.

The computed Magnetic Compton Profile (MCP) spectra for  Ni~\cite{re.ca.83,di.du.98,ku.as.90,ti.br.90,ba.zo.00,be.mi.12,ch.be.14,ch.be.14b,mi.eb.17} and Fe~\cite{co.co.89,ku.as.90,ta.sa.93,be.mi.12} have been previously reported in the literature. 
The analysis of spectra covers the aspects of multiple scattering, core contribution, 
relativistic effects and electronic correlations. 
The comparison with experimental measurements concerning the shape of the MCP spectra and the values of spin moments were also discussed. 
Along the [111] direction prominent features of the MCP for Fe and Ni are (i) the negative polarization of $s-$ and $p-$bands at low momentum, (ii) dips in the MCP profiles near $p_z=0 \  a.u.$, and the (iii) periodic features due to the Umklapp processes at momenta $\vec{p}=\vec{k} \pm n \vec{G}$, where $\vec{G}$ is the reciprocal-lattice vector and $n \in Z$. 
Generally, the theory overestimates the MCP spectra near $p_z=0$ irrespective of the band-structure method used in the Density Functional Theory (DFT)
calculations~\cite{re.ca.83,di.du.98,ku.as.90,ti.br.90,ba.zo.00,co.co.89,ku.as.90,ta.sa.93}. This discrepancy have been attributed to the
inadequate treatment of the electron-electron correlations in the Local Density Approximation (LDA) or its gradient corrected (GGA) type independent-particle-models for the exchange correlations of DFT.
It was shown during the last decades that Dynamical Mean Field Theory (DMFT)~\cite{MV89,GK96,KV04}, successfully removes some of the observed inconsistencies~\cite{LK01,CVA+03,MCP+05,GM07,GM12} in the description of the ground state properties of 3d transition metal elements. 
DMFT based calculations for the MCP profiles~\cite{be.mi.12,ch.be.14,ch.be.14b} of Fe and Ni, showed indeed that the low momentum discrepancies in the MCP are reduced, however high resolution measurements would be useful to investigate specific features in the MCP profiles that are still not well described. 

%

In this paper we report theoretical results on the MCP spectra for the disordered Fe$_{0.5}$Ni$_{0.5}$ alloy using the exchange correlation potential of LDA and the improved LDA+DMFT method~\cite{ko.sa.06}. We show that the discrepancies between the LDA and the experimental spectra at low momentum are corrected including local dynamic correlations captured by DMFT. At the same time we identify the correct magnitude of the Coulomb parameters on different alloy components. 
To study the interplay of disorder and correlation in momentum space, we compare MCP spectra of the Fe$_{0.5}$Ni$_{0.5}$ alloy with the corresponding spectra of an ordered FeNi alloy with simple cubic symmetry that has the same unit cell dimensions and chemical composition.  
We show that at the LDA level the total MCP spectra for both ordered and disordered FeNi alloys  are similar, while only within LDA+DMFT the distinction between the two becomes apparent. 
In particular at low momenta the MCP spectra obtained within the CPA calculation is reduced, while for the ordered alloy the MCP spectra is slightly enhanced. Based on the type decomposition of the MCP, we discuss differences and similarities between these spectra. As the MCP spectra of LDA+DMFT is found to be in agreement with the experiment in the disordered case, and as we are unaware of any previous experimental measurements or calculations of MCP spectra for the ordered FeNi alloy we thus, predict the MCP shape of FeNi along the [111] direction.

In the following section (Sec.~\ref{sec:det}) we provide an overview of the LDA+DMFT computational procedure of the MCP profiles within the Korringa-Kohn-Rostoker Green's function formalism for the disordered, Sec.\ref{sec:meth_dis}, and ordered, Sec.\ref{sec:meth_ord}, systems. The results are presented in section Sec.~\ref{sec:results}: the total MCP and type decompositions are discussed in Sec.~\ref{sec:mcp_tot_type_dis} and Sec.~\ref{sec:mcp_tot_type_ord} for the disordered respectively ordered alloys. The results for the density of states and magnetic moment calculations are presented in Sec.~\ref{sec:dos_mag}, and finally the paper is concluded in Sec.~\ref{sec:concl}.

\section{Computational details}
\label{sec:det}
The most important ingredient in the analysis of the electronic momentum density $n({\vec{p}})$ in disordered systems is the impurity configuration averaged Green's function. Such analysis was achieved for the first time by Mijnarends and Bansil~\cite{MB+76,MB+79} within the muffin-tin framework of the Coherent Potential Approximation, (CPA)~\cite{ve.ki.68,yo.mo.73,el.kr.74,ziman_79},
formulated using the multiple scattering theory, the so-called KKR-CPA~\cite{gyor.72,jo.ni.86,LSMS,vi.ab.01} method. Here we present results using the spin-polarized relativistic
Korringa-Kohn-Rostoker (SPR-KKR) method~\cite{E00,EKM11} in the atomic sphere approximation (ASA). The exchange-correlation
potentials parameterized by Vosko, Wilk and Nusair~\cite{VWN80}
were used for the LSDA calculations. For integration over the Brillouin zone the special points method has been used \cite{MP76}.
In addition to the LSDA calculations, a charge and self-energy self-consistent LSDA+DMFT scheme for correlated systems based on the KKR
approach~\cite{MCP+05,MM09,MI11,mi.eb.17} has been used. The many-body effects are described by means of dynamical mean field theory (DMFT)~\cite{MV89,GK96,KV04}
and the relativistic version of the so-called Spin-Polarized T-Matrix
Fluctuation Exchange approximation \cite{KL02, PKL05} impurity solver was used. The realistic multi-orbital interaction has been parameterized by the average screened Coulomb interaction $U$ and the Hund exchange interaction $J$. 
Recent developments allow to compute the dynamic electron-electron interaction matrix elements exactly~\cite{AIG+04}. It was shown that the static limit of the screened energy dependent Coulomb interaction leads to a U parameter in the energy range of 1 and 3 eV for all 3d transition metals~\cite{AIG+04}. As the $J$ parameter is hardly affected by screening it can be calculated directly within the LSDA and is approximately the same for all 3d elements, i.e J $\approx$ 0.9 eV. In our calculations we used values for the Coulomb parameter in the range of U = 2.0 to 3.0 eV and the Hund exchange-interaction J = 0.9 eV. The lattice parameter for both ordered and disordered alloy was taken as $6.763 \; a.u.$ and a BZ integration mesh of $ 62 \times 62 \times 62$ points was used.

The computation of Compton profiles within the SPR-KKR formalism~\cite{E00,EKM11} 
was worked out a decade ago~\cite{BME06,DB04}. 
The Magnetic Compton Profile is given by the momentum distribution of valence electrons projected along the scattering vector $p_z$.  
The spin projected momentum density is expressed in terms of the Green's function in the momentum representation, constructed from the real-space Green's function, using the eigenfunctions of the momentum operator. 
The electron momentum densities are usually calculated for the principal directions
$[001], [110], [111]$ using an rectangular grid of 200 points in each direction. The maximum value of the momentum in each direction is  $8 \; a.u.$. Here we present results only for the [111] direction which allows us to compare with experimental data~\cite{KKK+03} for the disordered alloy.

\subsection{Magnetic Compton profiles for disordered alloys, type decompositions}
\label{sec:meth_dis}

Given the eigenfunctions of the momentum operator, real-space integration is used to calculate $G_{m_s}(\vec{p},\vec{p},E)$. This integration is performed over a unit cell, and summed over the cells, as described in Ref.~\onlinecite{BME06}. 
In the momentum representation the ensemble averaged Green's function is given by:
\begin{widetext}
\begin{eqnarray}
G_{m_s}(\vec{p}, \vec{p}, E)
& = &
\frac{1}{\Omega}\,
\sum_{q} 
\sum_{A} x_{q_A}
\Big[
- \sum_{\Lambda} \tilde{M}^{q_A}_{m_s \Lambda m_s}
+\,
\sum_{\Lambda\Lambda '} M^{q_A}_{m_s \Lambda}
\Big(D^{q_A} \tau^{0q,0q}_{CPA}(E)\Big)_{\Lambda\Lambda '}
                         M^{q_A *}_{m_s \Lambda'}
\Big] \label{eq:1}
 \\
&+&\,
\frac{1}{\Omega}\,
\sum_{q} {\sum_{q'}}   e^{-i \vec{p} ( \vec{R}_{q} -  \vec{R}_{q'} )}
\sum_{A \ne B} x_{q_A}  x_{q^{\prime}_B}
\sum_{\Lambda \Lambda'}
 M^{q_A}_{m_s \Lambda}
\Big(D^{q_A} \tau^{nq,n^{\prime}q^{\prime}}_{CPA}(\vec{p},E)
 \tilde{D}^{q^{\prime}_B} \Big)_{\Lambda\Lambda '}
 M^{q^{\prime}_B *}_{m_s \Lambda'}\;. \nonumber
\end{eqnarray}
\end{widetext}

We denote by $q(q')$ the sites within the cells $n(n')$.
%
With $\tau^{0q,0q}_{CPA}(E)$ we denote the site-diagonal and with $\tau^{nq,n'q'}_{CPA}(\vec p,E)$ the site-non-diagonal, parts of the scattering path operator. 
In addition $\tau^{0q,0q}_{CPA}(E) =\int_{BZ} \tau_{CPA}(\vec{k},E) d^3 \vec{k}$.
The type-projected scattering path operators $D^{q_A}\tau^{0q,0q}_{CPA}(E)$ and $D^{q_A}\tau^{nq,n'q'}_{CPA}(\vec{p},E) \tilde{D}^{q'_B}$ appear as a consequence of the single-site approximation of the CPA, when computing the configuration average $\langle \tau^{nq,n^{\prime}q^{\prime}}_{\Lambda \Lambda^\prime} \rangle $~\cite{faul82,gonis_92}. 
Finally, $ M^{q,A}_{m_s \Lambda}$ and $\tilde{M}^{q,A}_{m_s \Lambda  m_s}$ are the regular and irregular Compton matrix elements for the alloy component A. 
The explicit form of these expressions was presented in Refs.~\onlinecite{BME06} and~\onlinecite{DB04}. 

In order to proceed with the decomposition of the MCP we shall in the following analyze Eq.~(\ref{eq:1}). A specific site $q$ in the unit cell contains the components A(B), with the concentrations $x_{q_{A(B)}}$. The site and component diagonal Green function in momentum representation is:
\begin{eqnarray}
G_{m_s}^{{A}, {A}}(\vec{p}, \vec{p}, E) \label{G_AA}
& = &
\frac{1}{\Omega}\,
\sum_{q} 
x_{q_A}
\Big[
- \sum_{\Lambda} \tilde{M}^{q_A}_{m_s \Lambda m_s}  \\
& +& \,
\sum_{\Lambda\Lambda '} M^{q_A}_{m_s \Lambda}
\Big(D^{q_A} \tau^{0q,0q}_{CPA}(E)\Big)_{\Lambda\Lambda '}
                         M^{q_A *}_{m_s \Lambda'} 
\Big] . \nonumber 
\end{eqnarray}
The site-diagonal but component-non-diagonal ($A \ne B$) Green function is obtained from the last term of Eq.~(\ref{eq:1}):
\begin{eqnarray}
G_{m_s}^{A,{B \ne A}}(\vec{p}, \vec{p}, E)
 = 
\frac{1}{\Omega}\,
\sum_{q} 
x_{q_A}  x_{q_B} \cdot  \label{G_AB} \\
\sum_{\Lambda \Lambda'} 
 M^{q_A}_{m_s \Lambda}
\Big(D^{q_A} \tau^{0q,0q}_{CPA}(\vec{p},E)
 \tilde{D}^{q_B} \Big)_{\Lambda\Lambda '} \nonumber
 M^{q_B *}_{m_s \Lambda'}
\end{eqnarray}

Accordingly, the spin resolved momentum densities in the disordered system are obtained integrating the corresponding Green's functions:
\begin{equation}\label{eq:np}
n_{m_s}^{A,B;X}(\vec p)={-\frac{1}{\pi}  \Im \int_{-\infty}^{E_F}
\left[  G_{m_s}^{A,B;X}(\vec p,\vec p,E) \right]  dE}\,
\end{equation}
With X we denote the functional form in the band structure calculation, X=LSDA(+DMFT) and $m_s=\uparrow(\downarrow)$. 
Using the expressions for the Green's function the pure Eq.~(\ref{G_AA}) and the  mixed Eq.~(\ref{G_AB})  
contributions in the momentum density can be obtained.
The double integral of the spin momentum density,  
projected onto the scattering direction ${\bf K}$, with $\vec{p}_z || {\bf K}$, defines the magnetic Compton profile (MCP):
\begin{equation}\label{mcsMCP}
J_{mag,\bf K}^{A,B; X}(p_z)=
\int \int [ n_{\uparrow}^{A,B;X}(\vec p) - n_{\downarrow}^{A,B;X}(\vec p) ] dp_x dp_y,
\end{equation}

\subsection{Site decomposition of the Magnetic Compton Profile for ordered alloys}
\label{sec:meth_ord}

For systems with more atoms in the unit cell, the MCP spectra is usually decomposed into the site-projected contributions and the interference-like terms similar to Ref.~\onlinecite{BME06}. 
The unit cell sites $q,q'$ can be occupied by atoms of type A or B.
The type- and site-diagonal Green's function has the form:
\begin{eqnarray}
G^{A,A}_{m_s }(\vec{p},\vec{p},E) 
&=&
  \frac{1}{\Omega} \sum_q
\Big[ 
\sum_{\Lambda\Lambda '} M^q_{m_s \Lambda}
\tau^{q,q}_{\Lambda\Lambda '}(\vec{p},E) M^{q *}_{m_s \Lambda'} \nonumber \\ 
 & - & 
 \sum_{\Lambda} \tilde{M}^{q}_{m_s \Lambda m_s}
\Big].  \label{G_qq} 
\end{eqnarray}
The summation over the sites ($q$) in Eq.~(\ref{G_qq}) is restricted to the sites occupied by the same type of atoms. 
The site-off-diagonal Green's functions is: 
\begin{eqnarray}
G^{A,B\neq A}_{m_s }(\vec{p},\vec{p},E) 
& = &
\frac{1}{\Omega} \sum_q \sum_{q' \ne q}
e^{-i\vec{p}\cdot(\vec{R}_q-\vec{R}_{q^\prime})}\cdot \nonumber \\
& & 
\sum_{\Lambda \Lambda'}
 M^{q}_{m_s \Lambda}
\tau^{q,q^\prime}(\vec p, E)_{\Lambda\Lambda '}
 M^{q^\prime *}_{m_s \Lambda'}
 \label{G_qq'}
\end{eqnarray}
Eq.~(\ref{G_qq'}) contains the product of Compton matrix elements $M^{q A}_{m_s \Lambda}$, $M^{q' B *}_{m_s \Lambda'}$ with the scattering path operator weighted by the phase factor  $\sum_{q, q'} e^{-i \vec{p} ( \vec{R}_{q} -  \vec{R}_{q'} )}$.  
In analogy with elementary formulas for X-ray diffraction by an assembly of atoms, equations of type Eq.~(\ref{G_qq'}) can be interpreted as interference or structure factor functions for the material. 
The momentum density and the magnetic Compton profile are then computed using the formulas Eq.~(\ref{eq:np}) and Eq.~(\ref{mcsMCP}). Accordingly, the magnetic Compton interference term is the MCP obtained using the Green's function of Eq.~(\ref{G_qq'}). 
Note that the Compton interference function can be an alloy-type non-diagonal or diagonal 
depending on the occupation of the $q(q')$-sites. 
This interference term is an incoherent scattering contribution and the corresponding MCP signal shall have a weak amplitude for all directions along $p_z$. 

\section{Results}
\label{sec:results}

\begin{figure}[h]
\includegraphics[width=0.45\linewidth, clip=true]{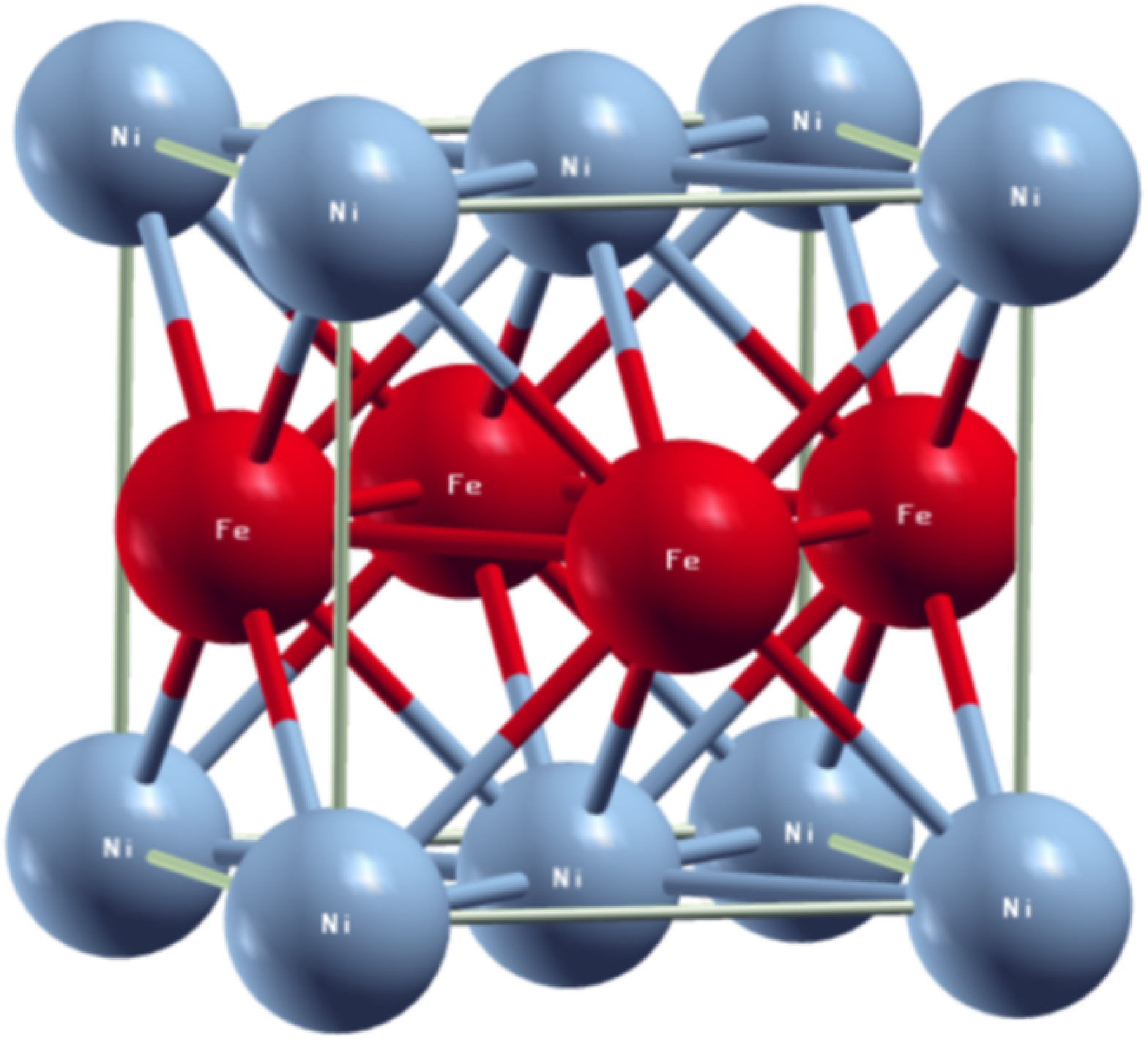}
\includegraphics[width=0.45\linewidth, clip=true]{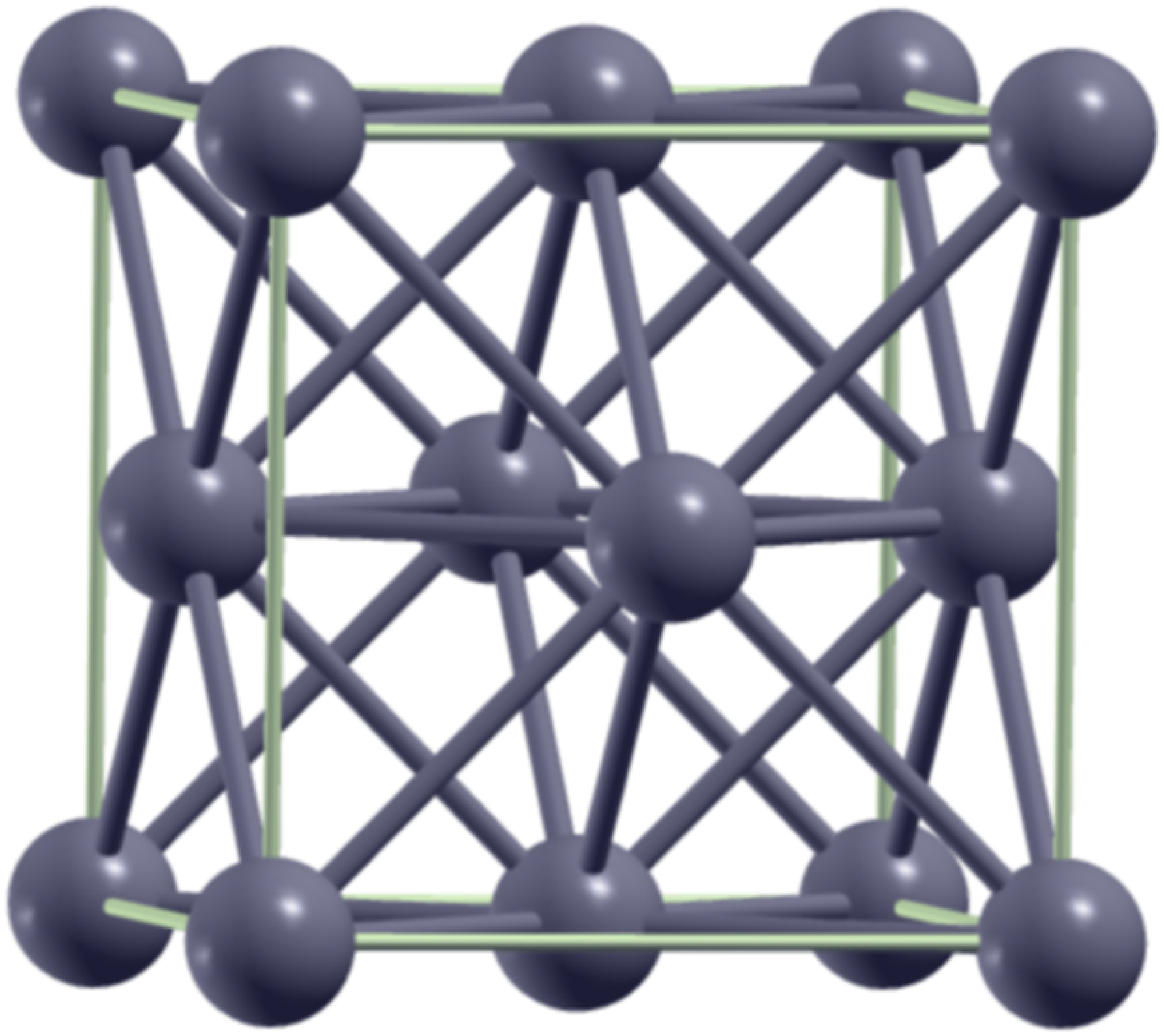}
\caption {\label{Fig:figure1}(Color on-line)
Left: The simple cubic structure of the ordered FeNi. Fe/Ni red/blue spehres. Right: the ``CPA effective'' atom (gray sphere) of composition Fe$_{0.5}$Ni$_{0.5}$.}
\end{figure}

In Fig.~\ref{Fig:figure1} we depict the unit cell for the ordered FeNi and the disordered Fe$_{0.5}$Ni$_{0.5}$ alloys in the simple cubic geometry. The ordered sc-FeNi alloy consists of alternating Fe and Ni layers with a unit cell containing two Fe and two Ni atoms.
%
%
In-plane atoms have neighbors of the same type, while out-of-plane neighbors are of different types. In this case, the calculation of the DMFT selfenergy is performed for each type separately, which allows to use different Coulomb/exchange parameters. 

In the fcc-geometry, within the CPA the ``effective''  Fe$_{0.5}$Ni$_{0.5}$ atom has the same neighbors in all directions. The DMFT (impurity) problem is still solved for each component Fe/Ni, in addition the CPA equation is imposed for self-consistency. The charge self-consistency involves the one ``effective'' atom unit cell.

\subsection{Total MCP and type decomposition spectra of Fe$_{0.5}$Ni$_{0.5}$ alloy}
\label{sec:mcp_tot_type_dis}

We have performed the LDA(+DMFT) calculations for the Fe$_{0.5}$Ni$_{0.5}$ alloy within CPA. Table~\ref{tab_I} summarizes the results for the spin and orbital magnetic moments. The DMFT calculations were done for different values of the local Coulomb interactions for Fe and Ni. As best values for the average Coulomb parameters we identified $U_{Fe}=2 eV$ and $U_{Ni}=3 eV$. The average exchange parameter was set to $J=0.9 eV$. 

\begin{figure}[h]
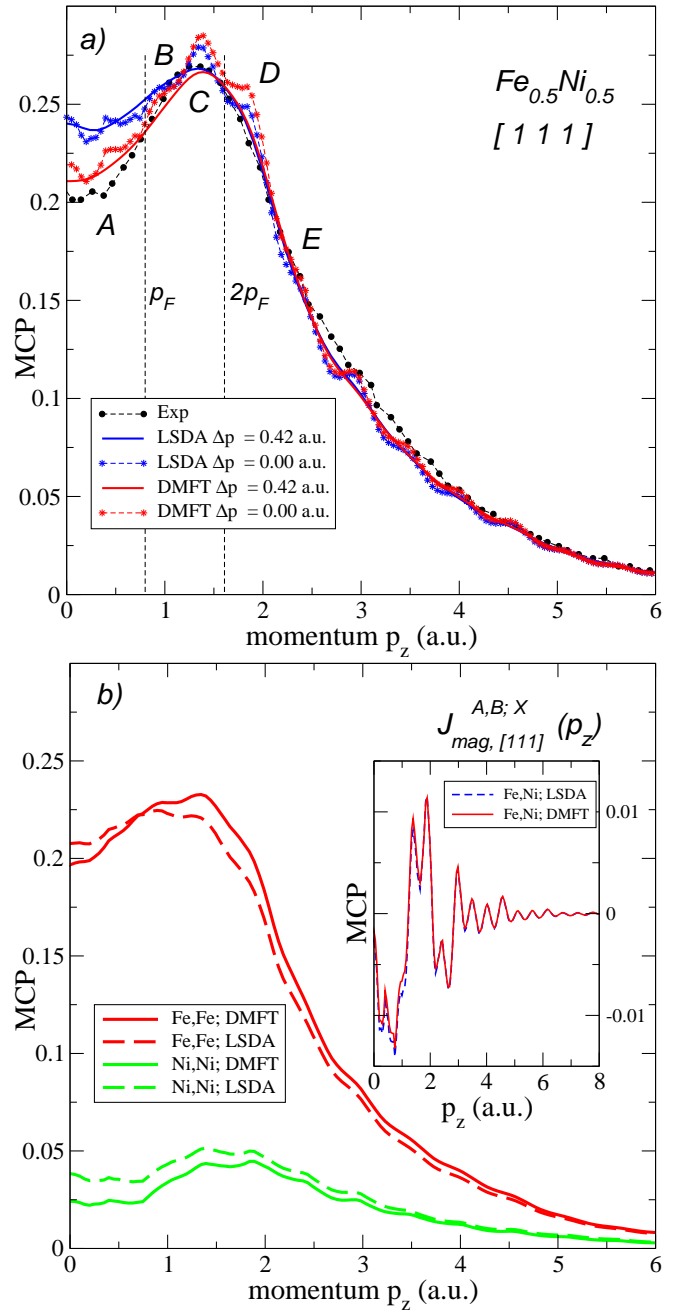

  \includegraphics[width=\linewidth, clip=true]{Fig2a.eps}
  \includegraphics[width=\linewidth, clip=true]{Fig2b.eps}
\caption {\label{Fig:figure2}(Color on-line)
a) Calculated total MCP of Fe$_{0.5}$Ni$_{0.5}$ alloy along [111] direction. Blue solid line: LSDA(CPA); red solid line: LSDA(CPA)+DMFT. The experimental spectra of Kakutani et al. \cite{KKK+03} (black circle). b) Type decomposition of MCP profiles of Fe$_{0.5}$Ni$_{0.5}$ alloy. LSDA/DMFT results are represented by dashed/solid lines. The inset shows the mixed term.
}
\end{figure}

The MCP spectra of Fe$_{0.5}$Ni$_{0.5}$ alloy are presented in Fig.\ref{Fig:figure2}. With decorated dashed blue/red lines we present raw LSDA/DMFT data. To allow the comparison with the experimental spectra of Kakutani et al.~\cite{KKK+03} the theoretical profiles have been broadened using a broadening parameter (full width at half maximum FWHM) equal with the experimental momentum resolution for recording the spectra, which was $\Delta p = 0.42\; \; a.u.$. These are seen in Fig.\ref{Fig:figure2}a) with solid blue/red lines. 

In the [111] direction one clearly observe the significant discrepancy between theory and experiment for $p_z < 1.5 \; a.u.$. The LDA+DMFT calculations capture the correct behavior at low momenta, similarly to the situation in bulk Fe and Ni~\cite{be.mi.12,ch.be.14,ch.be.14b}. Many of the specific features of the theoretical MCP can not be seen in the experimental profile due to the relative limited resolution ($\Delta p \approx 0.42 a.u.$).
Within the first zone, $p< p_F^{[111]} \approx 0.8 \ a.u $, the theoretical spectra predict a first peak marked with A and situated at $0.4 \ a.u$. This peak is absent in experiment but its Umklapp is observed experimentally (peak C). Within the second zone, the theory predicts peaks marked with B and C and outside the second zone the D and E peaks are visible. Further Umklapp features can be observed for larger momenta as shoulders at $\sim 2.9 a.u.$, $\sim 3.5 a.u.$ etc. Because of the relatively large broadening the experimental spectra ``melts'' the peaks B, C and D, therefore the Umklapp of A into the second Brillouin zone (C) is slightly overestimated. Furthermore the higher momenta Umklapp shoulders in the experimental profiles are considerably smeared out.

Based on Eq.~(\ref{mcsMCP}), the total MCP along the [111] direction has been decomposed, as seen in Fig.~\ref{Fig:figure2}b), into the type-projected contributions. Both $J_{mag}^{Fe Fe}$ and $J_{mag}^{Ni Ni}$ spectra show a pronounced dip at $p_z = 0$. At non-zero momenta we see that
electronic correlations leads to momentum density redistribution between different Brillouin zones. 
The $J_{mag}^{Fe Fe}$ DMFT spectra is situated below the LSDA spectra for $p_z < p_F^{[111]} \approx 0.8 \;a.u.$ and within the further Brillouin zones is above the LSDA. On the other hand the DMFT $J_{mag}^{Ni Ni}$ is situated below the LSDA spectra for the entire range of momenta. 
The inset of Fig.~\ref{Fig:figure2}b) shows the mixed MCP term, $J_{mag}^{Fe Ni}$ obtained from the formula Eq.~(\ref{eq:np}) and the Green's function Eq.~(\ref{G_AB}). The mixed term shows no significant correlation effects, its characteristic being the oscillatory structure. 

The type-resolved spectra has been scaled according to the spin moment obtained by self-consistent calculations, which is 1.57 $\mu_B$ in the LSDA calculations and 1.56 $\mu_B$ in the LSDA+DMFT calculations with U$_{Fe}$ = 2.0 eV and U$_{Ni}$ = 3.0 eV, respectively. 
Iron gives the dominant contribution in MCP, as a consequence of its large spin moment: 2.48~$\mu_B$ LSDA and 2.46~$\mu_B$  in LSDA+DMFT calculation, respectively. A significantly smaller Ni spin moment is obtained 0.66~$\mu_B$ (LSDA) and 0.65~$\mu_B$ (LSDA+DMFT) 
respectively. 
%
%


\subsection{Total MCP spectra and the site decompositions for the ordered FeNi alloy}
\label{sec:mcp_tot_type_ord}

\begin{figure}[h]
\includegraphics[width=\linewidth, clip=true]{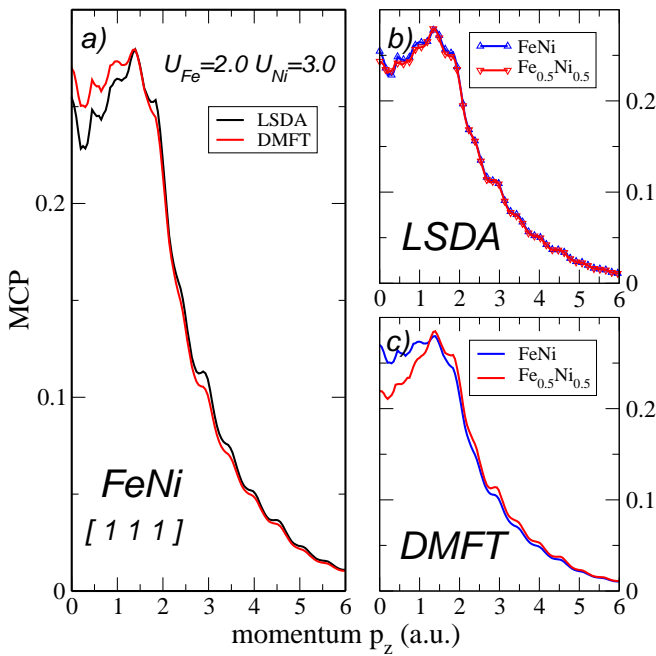}
\caption {\label{Fig:figure3}(Color on-line) Calculated MCP of the ordered FeNi alloy along [111] direction: panel a) the total MCP profiles computed with LSDA and DMFT for the values of $U_{Fe/Ni}=2/3eV$ and $J=0.9eV$. Panels b) and c) the total MCP profiles for the ordered and disordered FeNi alloys in LSDA and respectively in DMFT.}
\end{figure}

In Fig.~\ref{Fig:figure3}a) we present the comparison between the total MCP [111] profile of the ordered FeNi alloy obtained using the LSDA (black line) and the LSDA+DMFT (red line) methods. The computed spectra (no applied broadening) are normalized to the values of the magnetic moments obtained in LSDA/DMFT  calculations respectively.    

Contrary to the disordered case, Fig.~\ref{Fig:figure2}a) where correlation leads to a depleted spectra around the zero momenta, we predict that correlation effects enhance the MCP profile in the range up to $p_z \approx 1.5 a.u.$. For larger momenta $p_z > 1.5 a.u.$, similarly to the disordered case, the DMFT corrections does not change much on the LSDA shape.
We observe that at the LSDA level the MCP spectra can hardly distinguish between the ordered and disordered structures, Fig.~\ref{Fig:figure3}b). Any broadening applied to the spectra to account for the experimental resolution  would make the spectra identical. Including correlation effects Fig.~\ref{Fig:figure3}c) the LSDA(CPA)+DMFT spectra separate starting from the maximum value down to zero momenta and match the experimental results. The MCP spectra the red line Fig.~\ref{Fig:figure3}a) is our prediction for the shape of the MCP of ordered FeNi along the [111] direction. 

\begin{figure}[h]
\includegraphics[width=\linewidth, clip=true]{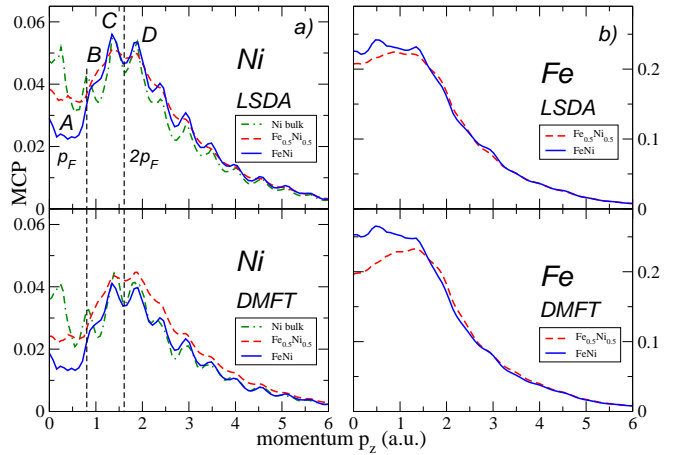}
\caption {\label{Fig:figure4}(Color on-line) Type decomposition of MCP spectra for the ordered FeNi alloy along [111] direction: panel a)/b) Ni respectively the Fe components computed within LSDA (upper part) and DMFT (lower part).}
\end{figure}

According to Eq.~(\ref{G_qq}) the total spectra is further decomposed into the MCP Ni Fig.~\ref{Fig:figure4}a) and MCP Fe Fig.~\ref{Fig:figure4}b) contributions. Note that Fe's weight to the total spectra is about four times larger than that of Ni. The reason why LSDA cannot distinguish between the ordered and disordered structure Fig.~\ref{Fig:figure3}b), become also apparent: the Ni/Fe MCP-components for the ordered alloy under/over estimate the corresponding spectra of the disordered alloy. The amount of under/over estimation nearly compensate each other producing a similar total spectra.  

The DMFT calculations for the disordered alloy produce slightly reduced spectra for Fe (red solid line) in comparison with the LSDA (red dashed line) as seen in Fig.~\ref{Fig:figure2}b) . For Ni a slightly stronger reduction takes place. 
For the ordered FeNi alloy, DMFT spectra of Fe/Ni are enhanced/diminished in the low momentum region, however, the increase of the Fe's MCP dominates the decrease on the Ni side, and an overall enhanced spectra is obtained  (Fig.~\ref{Fig:figure4}a; lower part).
Fig.~\ref{Fig:figure4}a, contains also the results of the Ni bulk MCP calculations. We mark the first essential peaks as in Fig.~\ref{Fig:figure2}a). Obvious differences are seen below $p_z < 2 p_F ~\approx 1.6 a.u.$ (first two BZ). For the bulk-Ni the first two peaks (A and B) are shifted and contained within the first BZ: the first (A) is positioned at about $0.23 a.u.$, the second (B) is in the vicinity of $p_F$. The third, the fourth and the subsequent Umklapp peaks are similar for all three spectra. 
The intensity of Umkalpps for Fe$_{0.5}$Ni$_{0.5}$ are smeared out additionally because of disorder effects. 
Electronic correlations enlarge the differences between the Ni-project spectra of the disordered alloy with respect to the bulk and ordered FeNi. Additionally the Ni contribution for the disordered alloy is further smeared out.

\subsection{Density of states and magnetic data analysis}
\label{sec:dos_mag}
\begin{figure}[h]
   \includegraphics[width=\linewidth, clip=true]{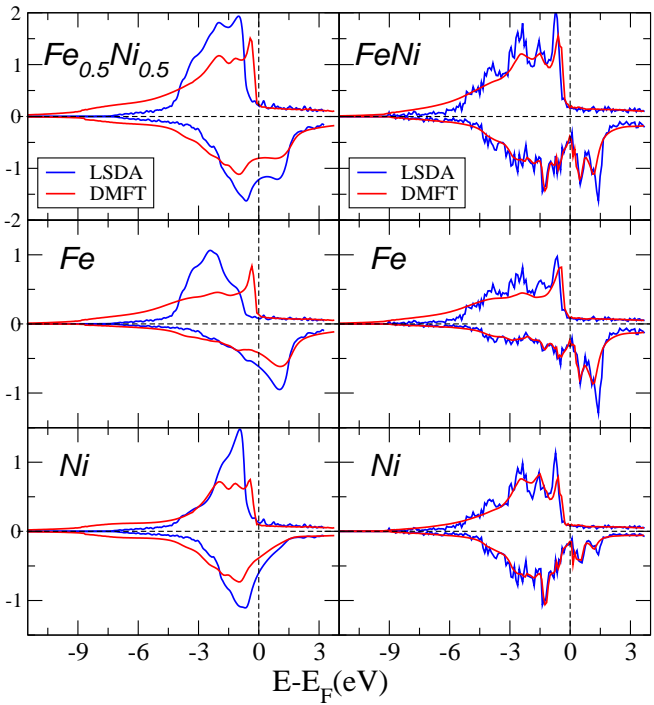}
\caption {\label{Fig:figure5}(Color on-line)
The density of state results for the disordered (left column) and ordered (right column) FeNi alloy. The DMFT results were obtained the values U$_{Fe}=2eV$, and U$_{Ni}=3eV$ and J=0.9eV.}
\end{figure}

In this section we discuss the ground state properties (DOS and magnetic moments) for the FeNi alloys. In Fig.~\ref{Fig:figure5}, we show the total and atom resolved DOS for Fe$_{0.5}$Ni$_{0.5}$ (left column) and FeNi alloys (right column). 
The combined effect of correlation and disorder is most remarkable for Fe's DOS, as seen also in the MCP spectra Fig.~\ref{Fig:figure2}. There is a strong renormalization of the spectral function towards the Fermi level on the majority spin channel (spin-up). For the minority spin channel the weight of DOS is suppressed. In both spin channels the spectra is broadened accordingly. 

Our calculations show that the majority spin channel of Ni undergoes relatively small changes upon addition of Fe. The minority states continue to remain occupied in the Fe$_{0.5}$Ni$_{0.5}$ alloy, as in Ni pure. The spectral changes are mainly limited to the weight reduction, in both spin channels, which is more significant than the spectral weight transfer towards the Fermi level, seen for the Ni majority spins. 

Comparing the cubic ordered with the fcc-disordered alloy, one can easily recognize that the LSDA(CPA)-DOS is more broadened because of the imaginary part of the complex effective potential. In addition electronic correlations have a less dramatic effect in the case of the ordered alloy, for the same values of the local Coulomb and exchange parameters.

Concerning the magnetic moments, the alloys have a ferromagnetic ground state. The Fe magnetic moment is in the range about $2.5 - 2.6 \mu_B$ while a value of about $0.6 \mu_B$ is obtained for Ni, depending on the strength of the local Coulomb parameters $U_{Fe/Ni}$. The Ni magnetic moment remains essentially at its value in bulk-fcc. The fact that Fe in FeNi has a larger moment than in fcc Fe at the same lattice constant can be explained by a smaller Fe-Ni hybridization due to the more contracted $3d$-orbitals of Ni. A larger hybridization tends to decrease the magnetic moment on an atom by filling the minority spin 3d orbitals, which are more extended.

\section{Conclusions}
\label{sec:concl}
The self-consistent spin polarized electronic structure and the Magnetic Compton profiles along the [111] direction have been computed for the  disordered Fe$_{0.5}$Ni$_{0.5}$ alloy. Disorder has been modeled using the CPA and the electronic correlations were considered through a multi-orbital Hubbard model solved with the DMFT. We showed that the discrepancy at low momenta due to the inadequate treatment of electronic correlations in LSDA can be corrected using DMFT. Note that DMFT has to be ``active'' on both alloy components. We have checked that neglecting ``electronic correlations'' on one of components or using improper values of the Coulomb interaction parameter does not provide a good comparison with the experimental spectra of Kakutani~\cite{KKK+03}. Most notably the LSDA(CPA)+DMFT with U$_{Fe}=2eV$, U$_{Ni}=3eV$ and J$=0.9eV$ resolves the discrepancy around $p_z=0$.  Umklapp features of the total-MCP spectra can be identified up to momenta $p_z < 2p_F$. 
Subject to electronic correlations, integrated spin-resolved momentum density show significant changes while integrated spin resolved real-space densities provide almost similar magnetic moments.
Due to the limited momentum resolution of the experimental spectra ($\Delta p =0.42 a.u.$) no clear comparison between theory and experiment can be performed for large $p_z$ momenta. High resolution measurements would be useful to identify specific features of the computed MCP profile. 
 
%


To study further the interplay between disorder and electronic correlations, in momentum space,  we have performed calculations for the MCP spectra of the ferromagnetic ordered FeNi alloy.
The calculation has been performed in the supercell setup with a similar lattice parameter as the one of the disordered alloy. Our results show that within the LSDA the MCP spectra for the ordered and disordered alloys are similar. This additional shortcoming of the LSDA can be explained by analyzing the type decomposition of MCP. Beyond the LSDA, the MCP spectra of ordered and disordered alloys are different, thus we predict within LSDA+DMFT the MCP shape of the FeNi alloy along the [111] direction.

An interesting conclusion maybe drawn from the results of the present work. 
Namely, MCP appears to be more sensitive to changes in the strength of the electronic correlations rather than in the different geometrical structures (different disorder realizations). In other words, electronic correlations affect the momentum distribution more significantly than the chemical bonding induced by structural disorder. 
From an experimental point of view high resolution measurements would be beneficial to resolve further the theoretical features of the MCP profile at low momentum and the blurring of the Umklapp features for high $p_z$.

\begin{widetext}

\begin{table}[h]
\begin{tabular}{cc|c|c|c|c|c}
\hline 
\multicolumn{2}{c|}{\multirow{2}{*}{$Fe_{0.5}Ni_{0.5}$}} & \multicolumn{4}{c|}{DMFT $U_{Fe/Ni}; J_{Fe/Ni}$ } & \multirow{2}{*}{LSDA} \\
\cline{3-6}
   &  & $2.0/0.0$ ; $0.9/0.0$ & $2.0/2.3$ ; $0.9/0.9$ & $0.0/2.0$ ; $0.0/0.9$ &   $2.0/3.0$ ; $0.9/0.9$ &  \\
\hline
\multirow{2}{*}{Fe} & $m_s(\mu_B)$ & 2.525 & 2.472  & 2.463 & 2.466 & 2.478  \\
                    & $m_l(\mu_B)$ & 0.089 & 0.10  & 0.059 & 0.105 & 0.058  \\ \hline
\multirow{2}{*}{Ni} & $m_s(\mu_B)$ & 0.628 & 0.653 & 0.680 & 0.653 & 0.659  \\
                    & $m_l(\mu_B)$ & 0.048 & 0.066 & 0.067 & 0.069 & 0.049  \\ \hline  
\hline 
\multicolumn{2}{c|}{\multirow{2}{*}{$FeNi$}} & \multicolumn{4}{c|}{DMFT $U_{Fe/Ni}; J_{Fe/Ni}$ } & \multirow{2}{*}{LSDA} \\
\cline{3-6}
   &  & $2.0/0.0$ ; $0.9/0.0$ & $2.0/2.3$ ; $0.9/0.9$ & $0.0/2.3$ ; $0.0/0.9$ &   $2.0/3.0$ ; $0.9/0.9$ &  \\
\hline
\multirow{2}{*}{Fe} & $m_s(\mu_B)$ & 2.594 & 2.596  & 2.572 & 2.599 & 2.573  \\
                   & $m_l(\mu_B)$ & 0.109 & 0.108  & 0.064 & 0.106 & 0.064  \\ \hline
\multirow{2}{*}{Ni} & $m_s(\mu_B)$ & 0.578 & 0.584 & 0.60 & 0.584 & 0.598  \\
                    & $m_l(\mu_B)$ & 0.037 & 0.040 & 0.04 & 0.045 & 0.051  \\ \hline
\end{tabular}
\caption{\label{tab_I} Magnetic moments: spin and orbital components for the for the disordered Fe$_{0.5}$Ni$_{0.5}$ and ordered FeNi alloy, computed with LSDA and for different values of $U_{Fe/Ni}; J_{Fe/Ni}$ using  LSDA+DMFT.}
\end{table}

\end{widetext}

\section{Acknowledgments}
We are indebted to Prof. D. Vollhardt for very useful discussions and for reading our manuscript.
Financial support of the Deutsche Forschungsgemeinschaft through FOR 1346, the CNCS - UEFISCDI (project number PN-II-RU-TE-2014-4-0009) and of the COST Action MP1306 EUSpec are gratefully acknowledged. JM would like to thank CEDAMNF project (CZ.02.1.01/0.0/0.0/15\_003/0000358) funded by the Ministry of Education, Youth and Sports of Czech Republic. 

\bibliography{paper}

\end{document}